\documentclass[sigconf]{acmart}
\usepackage{dsfont}
\usepackage{tabularx}
\usepackage{pdfpages}
\usepackage{amsmath,graphicx}

\AtBeginDocument{%
  }

\acmConference[MMSys'23]{the 31th ACM International Conference on Multimedia 
Systems, June 7--10, 2023, Vancouver, Canada}{June 7--10, 2023}{Vancouver, Canada}

\makeatletter
\newcommand\footnoteref[1]{\protected@xdef\@thefnmark{\ref{#1}}\@footnotemark}
\makeatother

\usepackage[normalem]{ulem} 

\copyrightyear{2023}
\acmYear{2023}
\setcopyright{rightsretained}
\acmConference[MMSys '23]{Proceedings of the 14th ACM Multimedia Systems
Conference}{June 7--10, 2023}{Vancouver, BC, Canada}
\acmBooktitle{Proceedings of the 14th ACM Multimedia Systems Conference
(MMSys '23), June 7--10, 2023, Vancouver, BC, Canada}
\acmDOI{10.1145/3587819.3590968}
\acmISBN{979-8-4007-0148-1/23/06}

\begin{document}

\title{Self-Supervised Contrastive Learning for Robust Audio -- Sheet~Music~Retrieval Systems}


\author{Lu{\'i}s Carvalho}
\authornote{Corresponding author.}
\affiliation{%
  \institution{Institute of Computational Perception}
  \city{Johannes Kepler University Linz}
  \country{Austria}}
\email{luis.carvalho@jku.at}

\author{Tobias Wash{\"u}ttl}
\affiliation{%
  \institution{Institute of Computational Perception}
  \city{Johannes Kepler University Linz}
  \country{Austria}}

\author{Gerhard Widmer}
\affiliation{%
  \institution{Institute of Computational Perception}
  \institution{LIT Artificial Intelligence Lab}
  \city{Johannes Kepler University Linz}
  \country{Austria}}

\renewcommand{\shortauthors}{Lu{\'i}s Carvalho et al.}

\begin{abstract}
Linking sheet music images to audio recordings 
remains a key problem for the development of efficient cross-modal 
music retrieval systems.
One of the fundamental approaches toward this task is to learn a cross-modal
embedding space via deep neural networks that is able to connect short snippets of 
audio and sheet music.
However, the scarcity of annotated data from real musical content affects the 
capability of such methods to generalize to real retrieval scenarios.
In this work, we investigate whether we can mitigate this limitation with 
self-supervised contrastive learning, by exposing 
a network to a large amount of real music data as a pre-training step, 
by contrasting 
randomly augmented views of snippets of both modalities, namely audio and sheet
images.
Through a number of experiments on synthetic and real piano data, we show that 
pre-trained models are able to retrieve snippets with better precision in all 
scenarios and pre-training configurations.
Encouraged by these results, we employ the snippet embeddings in the 
higher-level task of cross-modal piece identification and conduct more
experiments on several retrieval configurations.
In this task, we observe that the retrieval quality improves from 30\% up to 
100\% when real music data is present.
We then conclude by arguing for the potential of self-supervised contrastive 
learning for alleviating the annotated data scarcity in multi-modal music 
retrieval models.
Code and trained models are accessible at 
\url{https://github.com/luisfvc/ucasr}.
\end{abstract}




\maketitle

\section{Introduction}
\label{sec:intro}

Extensive amounts of music-related contents are available nowadays in 
the digital domain, in diverse forms, including studio and
live audio recordings, album covers, scanned sheet music, meta-data,
and video clips.
In addition, some of these contents are normally catalogued and/or curated with
manual effort by organizations from different contexts, 
such as cultural institutes, digital libraries, music publishers and 
concert halls.
Making such heterogeneous collections searchable and
explorable in an automated and content-based way  
requires powerful technologies for cross-linking between
items of different modalities.
As an example, a musician may have an incomplete excerpt of 
an unlabeled recording and wishes to retrieve from a digital 
database all relevant items in all possible modalities that 
are related to the query excerpt.

\begin{figure}[t!]
  \centering
  \includegraphics[width=\linewidth]{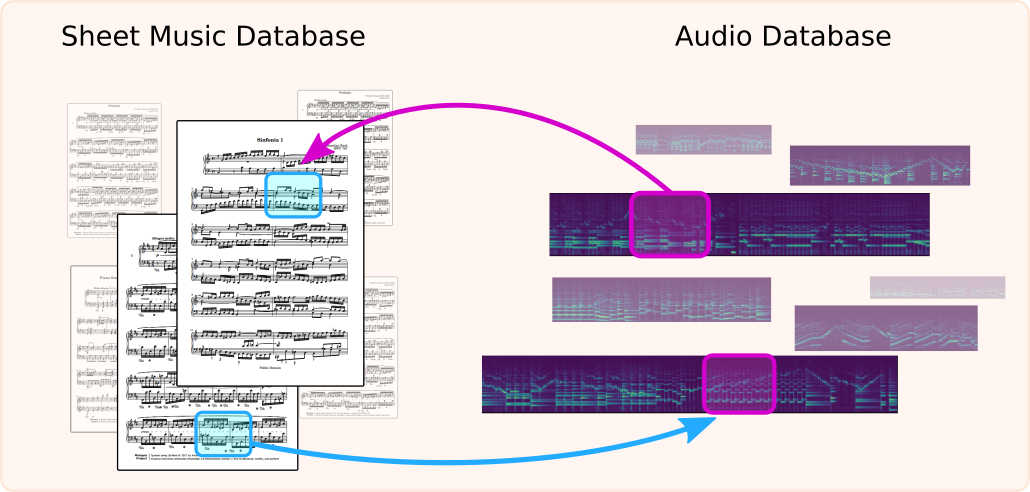}
  \caption{Illustration of the audio-sheet music snippet retrieval problem
  for both search directions. First, one wishes to query an audio excerpt 
  (on the right), represented by its magnitude spectrogram, and retrieve the 
  corresponding sheet music snippet from an image database (on the left).
  Analogously, one may wish to invert the search direction and retrieve 
  items from an audio database given a sheet music snippet input.
  All music visualizations were extracted from 
  the MSMD dataset~\cite{DorferHAFW18_MSMD_TISMIR}.}
\label{fig:teaser}
\end{figure}

A fundamental and challenging problem in many cross-modal music retrieval 
scenarios is referred to as 
\textit{audio--score retrieval}~\cite{MuellerABDW19_MusicRetrieval_IEEE-SPM}.
%
This problem is centered around two modalities, the acoustic one which 
comprises audio content, and the respective visual counterparts 
represented by music scores.
This task is defined as follows: given a query snippet in one modality (a short audio 
excerpt, for example), retrieve the corresponding item 
from a database in the other modality (a music score).
The most extreme and realistic setup for this task occurs when 
no metadata or machine-readable information of any kind (such 
as MusicXML or MIDI data) is available.
%
In this case, one is restricted to searching and retrieving only raw 
material, that is, audio recordings and digitized images of 
scanned sheet music.
With the aforementioned constraints, the underlying problem can be
referred to as
\textit{audio--sheet music retrieval}~\cite{DorferHAFW18_MSMD_TISMIR} 
and is illustrated in Figure~\ref{fig:teaser}.

Audio--score retrieval has been explored in numerous works 
in both intra- and inter-document scenarios and its applications 
are manifold.
%
The former scenario applies when both audio and score are known beforehand, and 
one wishes to obtain a fine-grained mapping between them.
This can be applied to score 
following~\cite{HenkelW21_RTScoreFollowing_Frontiers}, which is the 
real-time tracking 
of musical performances in the corresponding score, that can be employed 
in automatic sheet music page 
turning~\cite{ArztWD08_PageTurning_ECAI} for example.
Another use includes audio--score 
alignment~\cite{AgrawalWD21_CNNAttentionAlignment_IEEE-SPL}, 
where one wishes, 
for instance, to query a sequence of measures from a score and find the 
corresponding sections within an audio interpretation of the same 
piece~\cite{FremereyCME09_SheetMusicID_ISMIR}.

As for inter-document cases, the majority of scenarios address the 
task of 
\textit{audio--score piece identification}~\cite{BalkeALM16_BarlowRetrieval_ICASSP}, which can be defined both search directions: given an audio query, find a corresponding score from a collection; and given an excerpt from a score, retrieve an appropriate audio recording.
%
%
In the realm of modern digital music libraries, such technologies 
play an essential role in the indexing, navigation, browsing, and
synchronization of multi-modal 
databases~\cite{MuellerABDW19_MusicRetrieval_IEEE-SPM}.
One example of such applications is the Piano 
Music Companion~\cite{ArztBFFGLW14_PianoCompanion_PAIS}, 
a system that first tries to identify a piano piece that is 
being played, followed by synchonizing it within the corresponding 
score in real time.
However, a critical limitation of these systems is that they require 
the score to be available in a symbolic, machine-readable 
form -- e.g., MIDI or MusicXML -- which is a serious problem in 
practical applications.

Recent approaches for snippet-level audio--sheet music retrieval attempt 
to overcome this limitation by learning low-dimensional embeddings 
directly from the multi-modal data -- audio and scans or photographs of scores~\cite{DorferHAFW18_MSMD_TISMIR,
DorferAW17_ScoreIdentification_ISMIR,BalkeDCAW19_ASR_TempoInv_ISMIR}.
This is done by training a cross-modal deep convolution neural 
network (CNN) to project audio and score image snippets onto a shared space where 
semantically similar items of the two modalities will end up
close together, whereas dissimilar ones far apart.

Being of a fully supervised nature, this approach has a number 
of limitations.
First, it requires a large amount of labeled training 
data in order 
for a model to generalize to unseen data.
Second, such annotated data is of complex and expensive nature: 
it requires fine-grained alignments between time stamps on the 
audio signal and pixel coordinates 
in sheet music images in order to obtain matching cross-modal snippets.
The annotation process, besides being labor- and time-consuming, 
requires annotators with specialized musical training who are able to
correctly identify and interpret music notation in sheet music images 
and match them to note onsets in audio recordings.
For that reason current approaches rely solely on synthetic datasets, where both the scores and the audios -- and the corresponding annotations -- are generated from a symbolic score representation;
this results in poor generalization to real data, as we will 
demonstrate in our experiments (see Section~\ref{sec:exps}).

In this paper, we explore \textit{self-supervised contrastive learning} as 
way to mitigate the data scarcity problem in audio--sheet music snippet 
retrieval.
We propose to contrast
differently augmented versions of short fragments of audio recordings 
and sheet music images, as a pre-training step. The data for this task needs
no labels or annotations, so we have an almost infinite supply of this.
The key idea is that by trying to solve the pretext problem, the model 
can learn useful low-level representations, which can then be used for 
the audio--sheet music snippet 
retrieval task, where only few annotated data are available.

We conduct several experiments in datasets of different natures 
to demonstrate that the pre-training stage effectively alleviates 
the performance gap between synthetic and real data.
We then use the learned snippet embeddings for the downstream 
task of cross-modal \textit{piece identification} and observe 
improved retrieval performance in all models 
that were pre-trained. We summarize our contributions as follows.
\begin{itemize}
\item We design a method for multi-modal self-supervised contrastive 
    learning of audio--sheet music representations with publicly 
    available music data, where the network responsible for each modality 
    can be independently pre-trained and enabled for fine-tuning.
\item We show through detailed experiments on diverse data-sets
    that our models outperform the current state-of-the-art 
    method by a significant margin in the task of snippet retrieval.
\item As a proof of concept, we aggregate snippet embeddings to perform 
    cross-modal piece identification and demonstrate the effectiveness 
    of our improved models, which significantly outperform 
    fully supervised methods.
\end{itemize}

\section{Related Work}
\label{sec:review}

One of the key challenges in audio--sheet music retrieval refers to 
its multi-modality nature: finding some shared representation that allows
for an easy comparison between items from different modalities.
The traditional methods for connecting printed scores to their relative
audio recordings are based on common mid-level 
representations~\cite{MuellerABDW19_MusicRetrieval_IEEE-SPM,
IzmirliS12_PrintedMusicAudio_ISMIR}, such as 
chroma-based 
features~\cite{FremereyCME09_SheetMusicID_ISMIR,
BalkeALM16_BarlowRetrieval_ICASSP}, symbolic
representations~\cite{ArztBW12_SymbolicFingerprint_ISMIR}, 
or the bootleg 
score~\cite{YangTJST19_Sheet2MIDIRetrieval_Bootleg_ISMIR,
Tsai20_LinkingLakhtoIMSLP_Bootleg_ICASSP}, the latter being defined as 
a coarse codification of sequences of the main note-heads in a 
printed score.
However generating these mid-level representations involves  
pre-processing stages which are prone to error, such as optical 
music recognition~\cite{Calvo-ZaragozaHP21_OMRReview_ACM,
EelcoU_OMRAugs_ISMIR,LopezVCC21_OMRAug_ICDAR} and automatic music
transcription~\cite{BoeckS12_TranscriptionRecurrentNetwork_ICASSP,
SigtiaBD16_DNNPolyPianoTrans_TASLP}.

In order to avoid such unreliable pre-processing components, an 
alternative approach was proposed 
in~\cite{DorferAW17_ScoreIdentification_ISMIR,
DorferHAFW18_MSMD_TISMIR}, by designing a two-modal network that 
is able to learn a shared space between short fragments of score scans 
and their corresponding audio excerpts.
This is done by training the network to minimize the cosine distance 
between pairs of low-dimensional embeddings from snippets of audio 
and sheet music, and promising results on synthetic music data indicate 
the potential of replacing manually-designed common representations 
with learned spaces.
%





\section{The Proposed Method}
\label{sec:method}

In this section we first briefly describe how current 
approaches employ deep
CNNs to learn a cross-modal embedding space from 
pairs of matching audio and sheet music snippets.
Then we explain our proposed method, followed by 
describing the augmentation strategies for both 
sheet music and audio samples.

\subsection{Learning Audio--Sheet Music Embeddings}
\label{subsec:baseline}

The fundamental approach to learn correspondences between short snippets of 
music recordings and sheet music images was first proposed
in~\cite{DorferHAFW18_MSMD_TISMIR,DorferAW17_ScoreIdentification_ISMIR}.
This task is formulated as a cross-modal embedding learning 
problem, where a network is trained to optimize a shared space 
between the two modalities, by minimizing the cosine distance between
musically similar snippets whereas maximizing the distance between 
non-corresponding items.

\begin{figure}[tb]
  \centering
  \includegraphics[width=\linewidth]{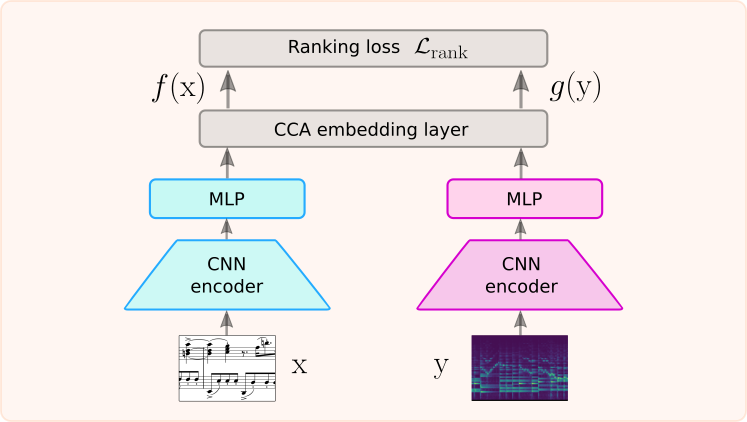}
  \caption{Illustration of audio--sheet retrieval model, adapted 
  from~\cite{DorferHAFW18_MSMD_TISMIR}.
  The left and right independent pathways encode sheet 
  music and audio snippets, respectively, by projecting together
  corresponding cross-modal pairs while maximizing the distance 
  between non-corresponding ones.}
\label{fig:bl_schema}
\end{figure}

The network, which is illustrated in Figure~\ref{fig:bl_schema}, 
consists of two independent pathways, each responsible for one 
modality.
Each pathway is composed of a VGG-style 
encoder~\cite{SimonyanZ14_VGGStyle_ICLR}, followed by 
a multi-layer perceptron layer (MLP) that learns higher-level 
abstractions from the encoder output.
At the top of the network a canonically correlated (CCA) embedding 
layer~\cite{DorferSVKW18_CCALayer_IJMIR} is placed, forcing the two 
pathways to learn representations that can be projected onto 
a 32-dimensional shared space.

Then a pairwise ranking loss~\cite{KirosSZ14_VisualSemanticEmbeddings_arxiv} 
is employed to minimize the distance between 
embeddings from matching snippets of different modalities.
Let $(\mathbf{x},\mathbf{y})$ represent a pair of corresponding 
sheet music and audio snippets (positive pairs), 
as displayed in Figure~\ref{fig:bl_schema}.
The sheet music pathway is represented by the function $f$, while $g$ 
denotes the audio embedding function.
The functions $f$ and $g$ map $\mathbf{x}$ and $\mathbf{y}$ to the 
shared low-dimensional space.
Then the similarity function $\mathrm{sim}(\cdot)$, 
defined as the cosine similarity, 
is used to compute the final ranking loss :
\begin{equation}
\mathcal{L}_{\mathrm{rank}} = 
\sum_{(\mathbf{x},\mathbf{y})} \sum_{k=1}^K
\mathrm{max} \Bigl\{
0, \alpha - 
\mathrm{sim} \Bigl( f(\mathbf{x}),g(\mathbf{y}) \Bigl) + 
\mathrm{sim} \Bigl( f(\mathbf{x}),g(\mathbf{y}_k) \Bigl)
\Bigl\} \mathrm{,}
\end{equation}\label{eq:rank_loss}
where $\mathbf{y}_k$ for $ k \in {1, 2, \ldots, K} $ represent additional 
contrastive (negative) examples from $K$ non-matching snippets within the same 
training mini-batch.
This ranking loss is applied on all $(\mathbf{x},\mathbf{y})$ pairs of each 
mini-batch iteration, and the margin parameter $\alpha \in \mathbb{R}_{+}$ in 
combination with the $\mathrm{max}\left \{ \cdot \right \}$ function 
penalize matching snippets that were poorly embedded.

After the training is done, the snippet retrieval task illustrated in 
Figure~\ref{fig:teaser} can then be easily and efficiently performed 
via nearest-neighbor search in the shared space.

\subsection{Self-Supervised Contrastive Learning}
\label{subsec:main_method}

\begin{figure}[tb]
  \centering
  \includegraphics[width=\linewidth]{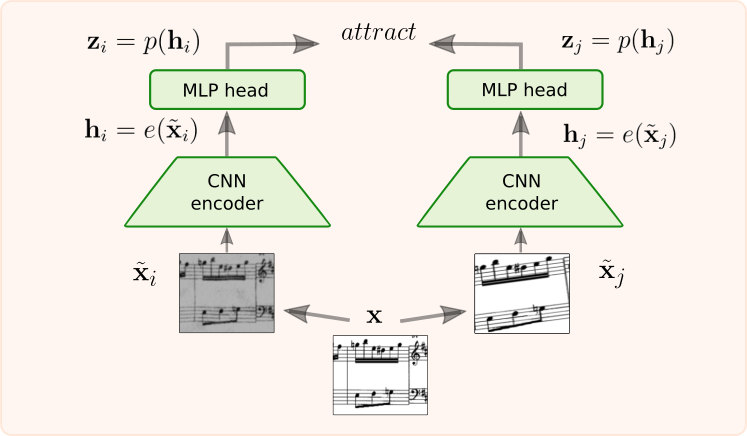}
  \caption{Sketch of our proposed self-supervised contrastive
  learning architecture, adapted from~\cite{ChenKNH20_SimCLR_ICML}, 
  for an example of sheet music snippet.
  Two different views are rendered using augmentation transforms 
  (contrast and rotation, for example), which are fed to a CNN 
  encoder followed by a MLP head, generating a positive pair of
  embeddings $(\mathbf{z}_i, \mathbf{z}_j)$, which should be 
  projected close together.  
  }
\label{fig:SimCLR}
\end{figure}

In this work we build on the SimCLR 
framework~\cite{ChenKNH20_SimCLR_ICML}, 
a self-super-vised contrastive method for image representations.
The goal is to learn useful representations from unlabeled data 
using self-supervision .
The idea is to train a network encoder to be as invariant 
as possible concerning a set of given augmentation 
transforms~\cite{DosovitskiySRB14_InstanceDiscrimination_NIPS}.
In order to do that, different augmentations are applied to a training 
sample so two distinct views thereof are generated (which constitute a "positive pair" that represent the same item).
Then a Siamese network~\cite{ChopraHL05_SiameseNNs_CVPR} 
encodes both views into embeddings, and a 
contrastive loss function is applied in order to bring  together 
latent representations from the same sample, while pushing away 
embeddings of negative pairs.

This approach is sketched in Figure~\ref{fig:SimCLR} for the case of sheet 
image snippets, however we stress the procedure is analogous for the audio 
case. More precisely, the following steps are performed:
\begin{itemize}
\item Given a sample $\mathbf{x}$ from the training mini-batch, two 
    stochastic sets of data augmentation transforms are applied
    to $\mathbf{x}$ to render two different augmented views of the 
    same sample (a ``positive pair"), 
    namely $\tilde{\mathbf{x}}_i$ and $\tilde{\mathbf{x}}_j$.
    (Our specific data augmentation pipeline for each modality is described in Section~\ref{sec:data_augs} below.)
\item Then a CNN encoder $e(\cdot)$ is used to compute a latent 
    representation $ \mathbf{h}_i = e(\tilde{\mathbf{x}}_i) $ for each
    view.
\item An MLP projection head $p(\cdot)$ maps the 
    encoder latent embedding $\mathbf{h}_i$ to a final space 
    $\mathbf{z}_i = p(\mathbf{h}_i) $ where 
    the contrastive loss is used.
\item Then the normalized-temperature cross-entropy
    ($NT-Xent$) loss 
    function~\cite{Sohn_NPairObjective_NeurIPS}
    is applied and summed over all positive augmented pairs 
    $(\tilde{\mathbf{x}}_i,\tilde{\mathbf{x}}_j)$ within 
    the mini-batch:
\begin{equation}
    \mathcal{L} = 
    \sum_{(i,j)} \mathrm{log}
    \frac{\mathrm{exp}(\mathrm{sim}(z_i, z_j) / \tau)}
    {\sum_{v=1}^{2N}\mathds{1}_{[v \neq i]}
    \mathrm{exp}(\mathrm{sim}(z_i, z_v) / \tau)} \mathrm{.}
\end{equation}\label{eq:nt_xent}
\end{itemize}

During training, two different 
augmented views are rendered from each sample within the 
mini-batch of size $N$, yielding a pool of $2N$ augmented 
views per mini-batch, over which the above loss function is
applied.
For each positive pair, all remaining $2N-2$ samples within the 
mini-batch are considered negative samples, as indicated by the 
summation in the denominator of Equation~(2). 
The temperature parameter $\tau \in \mathbb{R}_{+}$ works 
similarly to the margin parameter $\alpha$ in 
Equation~(1), prioritizing poorly 
embedded samples.

In this architecture the MLP projection layer $p$ is 
employed only during self-supervised learning.
After the model is trained, this layer is discarded and only 
the encoder $e$ is used as a pre-trained model for a given 
downstream task, which in our case is audio--sheet music
retrieval.
As discussed in~\cite{ChenKNH20_SimCLR_ICML}, the reason 
is that empirical results show that applying the 
contrastive loss over a function $p$ of the encoder 
embeddings 
$\mathbf{z}_i = p(\mathbf{h}_i) $
during training is beneficial because it improves the 
quality of learned representations.

An important difference between our approach and the method described 
in ~\cite{ChenKNH20_SimCLR_ICML} is that in
our setup we have two separate convolutional pathways, one 
responsible for encoding each modality (see Figure~\ref{fig:bl_schema}).
We perform self-supervised contrastive 
learning separately in each of the modalities, in order to 
obtain two separate and independent pre-trained encoders.
Since the pathways for audio and sheet music are independent,
we can simply select the modality we wish to pre-train, 
and obtain a pre-trained
encoder for the given modality.
The encoder is then placed in the multi-modal network 
in Figure~\ref{fig:bl_schema} and fine-tuned for the
audio--sheet music retrieval task.

Our CNN encoder follow the setup in~\cite{DorferHAFW18_MSMD_TISMIR}.
The encoder architecture is the same in each modality, 
and consists of a 
VGG-style network~\cite{SimonyanZ14_VGGStyle_ICLR} with eight 
convolutional layers, each of them followed by a 
batch normalization layer~\cite{IoffeS15_BatchNorm_ICML} 
and exponential linear unit (ELU)~\cite{ClevertUH16_ELU_ICLR}
activation.
A max pooling layer is applied every two consecutive 
convolutional layers in order to halve the dimensions 
of the hidden representations.

Our projection head $p$ consists of an MLP with one 
hidden layer followed by batch normalization and rectified 
linear unit activation (ReLU)~\cite{Agarap18_ReLU_CVPR}, 
from which the output embedding
is L2-normalized and mapped to a 32-dimensional final 
representation, on which the contrastive loss is calculated.
%

\section{Data Augmentations}
\label{sec:data_augs}

In self-supervised learning, one wishes to 
optimize a model so it can be highly invariant in 
regards to a set of augmentation transforms.
Therefore a proper composition 
of data augmentation operations is crucial for learning 
good representations~\cite{ChenKNH20_SimCLR_ICML}.
In our system, an augmented view $\tilde{\mathbf{x}}_i$ is generated 
by applying a pipeline of $M$ augmentation transforms on the 
original sample $\mathbf{x}$.
Each augmentation transform $t_m()$ has an independent 
probability $p_m$ to be applied to $\mathbf{x}$.
Each time the transform $t_m()$ is selected, its 
hyper-parameters are stochastically sampled from 
a pre-defined distribution, which is particular for each 
transform.

In the following we provide details of the augmentations 
we employed during the self-supervised training of 
each modality, as well as information about the used 
datasets.

\subsection{Sheet Music Augmentation Transforms}

\begin{figure}[tb]
  \centering
  \includegraphics[width=\linewidth]{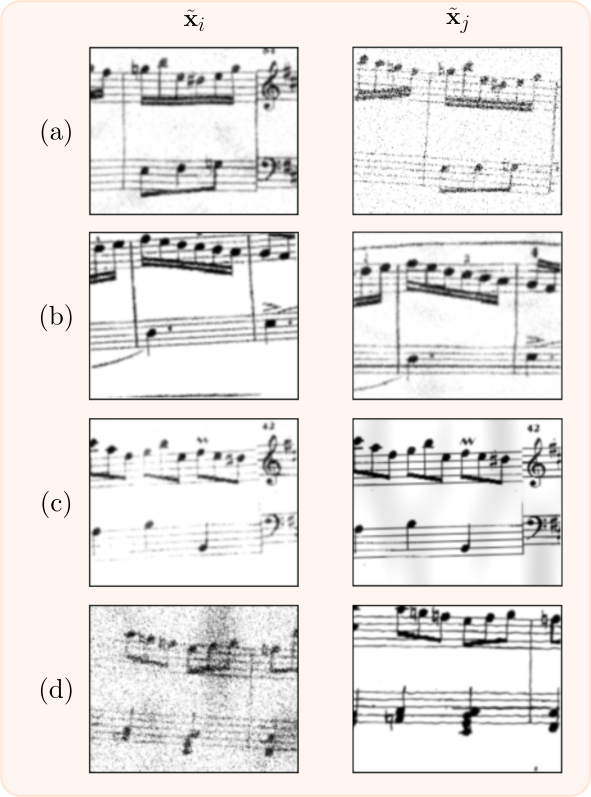}
  \caption{Examples of four pairs of augmented sheet music 
  snippets after all nine transforms were randomly applied.
  One should note that, even though the excerpts were greatly
  corrupted, they are still readable.
  These examples were obtained from 
  the MSMD dataset~\cite{DorferHAFW18_MSMD_TISMIR}.}
\label{fig:score_augs}
\end{figure}

Augmentation strategies have proven to be powerful techniques
to help machine learning models generalize to unseen data in 
image 
tasks~\cite{KrizhevskySH12_ImageNet_NIPS,ShortenK19_DataAugSurvey_JBD}.
In sheet music analysis, augmentation transforms are chosen so that
they can emulate document variations and degradations 
of various types~\cite{EelcoU_OMRAugs_ISMIR,
LopezVCC21_OMRAug_ICDAR,Calvo-ZaragozaHP21_OMRReview_ACM}.
We build on these works and define a set of nine transforms that
are applied to the sheet music snippets, which 
are described as follows.
\begin{itemize}
    \item We shift the snippet horizontally (1) and vertically (2) in
    relation to its positive pair.
    The horizontal shift is calculated in a way that positive pairs will
    share at least 80\% of their content, and 75\% 
    for the vertical shift.
    \item The snippet is resized (3) to have between 90 and 110\% of
    its original size.
    \item The snippet is rotated (4) to a maximum 
    angle of 8 degrees, counter- or clockwise.
    \item We apply Additive White Gaussian Noise (AWGN) (5) and 
    Gaussian Blur (GB) (6), to simulate noisy 
    documents and poor resolution, respectively.
    \item Additive Perlin Noise (APN) (7)~\cite{Perlin02_PerlinNoise_ACMTOG} 
    is added to the sample.
    This transform generates big darker and lighter areas in the
    image, mimicking quality differences in the image snippet.
    \item Then random small (8) and large (9) elastic 
    deformations (ED)~\cite{SimardSP03_ElasticTransfor_ICDAR}
    are applied, generating wave-like distortions to the 
    image, whose strength and smoothing can be tuned.
    Small EDs are applied on small scales, with the effect of
    deforming the shapes of smaller symbols and lines.
    When large EDs are applied, the structure and orientation 
    of bigger music symbols are modified, for example by skewing or bending 
    bar lines and note symbols, and squeezing or elongating 
    note heads.
\end{itemize}
The augmentations are applied in the presented order and
we tune the hyper-parameters of each individual transform 
in a way that a snippet is highly degraded, but still legible.
Figure~\ref{fig:score_augs} shows four examples of augmented snippet 
pairs when all nine transforms are stochastically applied to four 
sheet music snippets.

\subsection{Audio Augmentation Transforms}

Several works have successfully explored data augmentation 
for several audio and music learning 
tasks~\cite{ParkCZCZCL19_SpecAugment_INTERSPEECH,
SalamonB17_CNNDataAugForEnvSoundClass_IEEE-LettersSP,
SchlueterG15_DataAug_ISMIR,TakahashiGPG16_CNNDataAugForEventRecog_INTERSPEECH}.
We build on them and in the following define the sequence 
of eight audio transforms used to augment 
audio excerpts.
\begin{itemize}
    \item We apply a time shift (1) between the two excerpts of
    a positive pair.
    The shift is calculated in a way that corresponding snippets
    will share at least 70\% of their content.
    \item Polarity inversion (2) is applied to the audio excerpt
    by multiplying its amplitude by $-1$.
    \item Additive White Gaussian Noise (3) with a signal-to-noise
    ratio between 5 and 20 dB is added.
    \item A gain change (4) between $-12$ and $12$ dB is applied 
    to the signal.
    \item We apply a seven-band parametric equalizer (5) in order to 
    adjust the volume of seven different randomly-chosen frequency 
    bands.~\footnote{\url{https://iver56.github.io/audiomentations/waveform_transforms/seven_band_parametric_eq/}}
    \item The audio excerpt is stretched in time (6) without 
    modifying its pitch by changing the tempo with a random 
    factor between 0.5 and 1.8.
    \item Time (7) and frequency (8) masks are applying to the 
    audio snippet \textit{{\`a} la} 
    SpecAugment~\cite{ParkCZCZCL19_SpecAugment_INTERSPEECH}.
    Both time and frequency largest masks correspond to $20 \%$
    of the snippet duration and frequency range, respectively.
\end{itemize}
The augmentations are applied in the order they were declared
above.
The transforms 1-5 are applied directly on the waveform 
snippets, while transforms 6-8 are applied
in the frequency domain due to computational benefits.

\section{Experiments and Results}
\label{sec:exps}

In this section, we report on the experiments conducted 
to validate our proposed method.
We first briefly elaborate on the pre-processing steps, 
dedicated datasets and training setup.
Then we carry out experiments on cross-modal snippet 
retrieval and piece identification.

\subsection{Snippet Preparation}
\label{subsec:snippet_prepare}
In the following, we describe how the snippets are 
extracted, pre-processed and prepared 
for training.

\subsubsection{Sheet Music Snippets}
Our sheet music images are first re-scaled to a $1181 \times 835$ 
resolution (pixels) per page.
Then $160 \times 200$ snippets are selected in such a way that they comprise 
musical content, \textit{i.e.} within the systems of the document 
(groups of two staves, for piano sheet music).
When no annotation is available
concerning 
pixel coordinates of note heads and/or system locations 
(\textit{i.e.}, in the raw data for self-supervised learning), we 
use the Audiveris 
engine~\footnote{\url{https://audiveris.github.io/audiveris/}} 
to automatically detect the staff lines as a pre-processing stage.
Manual inspections showed that Audiveris is able to properly 
identify system coordinates in printed piano scores with accuracy 
of over $99\%$, therefore it is unlikely that snippets will not 
exhibit musical content.
Examples of sheet music snippets are depicted in 
Figures~\ref{fig:bl_schema}, \ref{fig:SimCLR} 
and~\ref{fig:score_augs}

\subsubsection{Audio Snippets}
Our music datasets consist of audio signals with a sampling rate
of $22.05$ kHz.
The log-frequency spectrogram of each signal is computed with a resolution
of $20$ frames per second and minimum and maximum frequencies of $30$~Hz 
and $6$~kHz respectively, generating $92$~frequency bins per frame.
We then cut out 84 frames of audio (approximately 4.2 seconds) to generate 
a snippet, which has a final shape of $92 \times 84 $ (bins~$\times$~frames).
Examples of audio log-spectrograms and snippets are shown in 
Figures~\ref{fig:teaser} and~\ref{fig:bl_schema}.

\subsection{Datasets}
\label{subsec:datasets}

To pre-train the sheet music encoder, we scrape 
data from the International Music Score Library Project 
(IMSLP)~\footnote{\url{https://imslp.org/wiki/Main_Page}}, 
an online plataform that hosts public domain music scores.
We collect 3,485 scanned piano scores relating to 842 music pieces
by several composers, which amounts to approximately 7,000 
pages of sheet music.
From these documents we extract over 700k snippets as described 
in~\ref{subsec:snippet_prepare} for training and validation.
We will provide the IMSLP links to all music scores of our collection 
in the paper 
repository~\footnote{\label{fn:repo}\url{https://github.com/blinded_for_review}.}.

For self-supervised learning of the audio encoder, 
we use the recordings from MAESTRO~\cite{HawthorneSR+_MAESTRO_ICLR}, 
a public dataset with 1,276 piano recordings comprising around 
200 hours of piano music.
Since there is no test stage at pre-training, we merge the pre-defined MAESTRO test 
split into the train set, and generate around 840k audio snippets 
as described 
in~\ref{subsec:snippet_prepare} to train and validate the audio encoder.

To train the final audio--sheet music network, we use the 
Multi-Modal Sheet Music Dataset (MSMD)~\cite{DorferHAFW18_MSMD_TISMIR},
which is a database of polyphonic piano music and scores.
With over 400 pieces covering over 15 hours of audio, this dataset has 
fine-grained cross-modal alignments between audio note onsets and 
sheet music note-head coordinates, which makes it suitable for 
generating matching audio--sheet music snippets.
This dataset is of fully artificial nature: audio recordings are 
synthesized from MIDI files using 
FluidSynth~\footnote{\url{https://www.fluidsynth.org/}} and the 
scores are engraved with 
LilyPond~\footnote{\url{http://lilypond.org/}}.
The matching snippets are extracted in a way that they are centred around 
the same note event, being the note onset for the audio side and the note-head 
pixel coordinate for the sheet music side.

In our experiments, we wish to investigate how well pre-training helps to generalize from synthetic to real data.
To this end, we evaluate on three datasets: on a (1) fully 
artificial one, and on datasets consisting (2) partially 
and (3) entirely of real data.
For (1) we use the test split of MSMD and for (2) and (3) we 
combine the Zeilinger and Magaloff
Corpora~\cite{CancinoChaconGWG_Magaloff_ML} 
with a collection of commercial recordings 
and scanned scores that we have access.
These data account for more than a thousand pages of sheet music scans 
with fine-grained mappings to both MIDI files and over 20 hours 
of classical piano recordings.
We then define the following evaluation sets.
(2) \textit{RealScores\_Synth}: a partially real set, with \textit{scanned} (real) 
scores of around 300 pieces aligned to notes of \textit{synthesized} MIDI 
recordings.
And (3) \textit{RealScores\_Rec}: an entirely real set, with \textit{scanned} (real) 
scores of around 200 pieces with fine-grained alignments to 
\textit{real audio} recordings.

\subsection{Training Setup}
\label{subsec:setup}

Our learning pipeline is split into two stages: (i) self-supervised
learning on each individual modality with a batch size of 256, 
followed by (ii) cross-modal training on pre-loaded encoders from 
either or both modalities, with a batch size of 128 pairs, where 
audio and sheet music snippets are project onto a 32-dimensional 
space.

In both stages we use the Adam optimizer~\cite{KingmaB15_Adam_ICLR} 
and He initialization~\cite{HeZRS15_HeInit_ICCV} in 
all convolutional layers.
The temperature parameter $\tau$ and triplet margin $\alpha$ are set to 
0.5 and 0.6, respectively.
We set the initial learning rates of (i) and (ii) to 0.001 and 0.0004 
respectively.
We observe the validation loss during training and halve the learning
rate if there are no improvements over 10 consecutive epochs, apply
early stopping when halving happens five times, and select the best
model among all epochs for testing.
For sake of simplicity, we leave the remaining details concerning 
topological design of the networks, further learning hyper-parameters, 
and augmentation probabilities and hyper-parameters, to our
repository.~\footnoteref{fn:repo}

\subsection{Snippet Retrieval Experiments}
\label{subsec:exp_1}

In this section we evaluate a two-way snippet retrieval task: 
given a query excerpt, retrieve the corresponding 
snippet in the other modality.
This is done by first embedding the query excerpt and all snippets of
the target modality, and then selecting the query's nearest 
neighbor in the embedding space as the best match, based on 
their pairwise cosine distance.

For each of the three evaluation datasets introduced 
in section~\ref{subsec:datasets}, we select a pool of 
10,000 audio--sheet music snippet pairs for evaluation.
We perform the retrieval task in both search directions: 
audio-to-sheet music (A2S) and sheet music-to-audio (S2A).

As evaluation metrics we compute the 
\textit{Recall@k} (R@k), \textit{Mean Reciprocal Rank} (MRR) and
the \textit{Median Rank} (MR) for each experiment.
The R@k measures the ratio of queries which were correctly 
retrieved within the top $k$ results.
The MRR is defined as the average value of the reciprocal 
rank over all queries, with the rank being the position 
of the correct match in the distance-ordered ranked list of 
candidates. MR is the median position of the correct match 
in the ranked list.

We perform snippet retrieval with the 
state-of-the-art 
method~\cite{DorferHAFW18_MSMD_TISMIR}, which will be
denoted as the baseline \textit{BL}, and compare with 
all possible combinations of self-supervised pre-training 
as we proposed.
Since in the cross-modal network the two convolutional 
pathways responsible for encoding each modality are 
independent, we can load either or both encoders 
with parameters that were pre-learned before 
training.
We then define the following  models:
\begin{itemize}
    \item \textit{BL+A}: the audio encoder is pre-trained
    \item \textit{BL+S}: the sheet music encoder is pre-trained
    \item \textit{BL+A+S}: both audio and sheet music encoders 
    are pre-trained,
\end{itemize}
which are modified versions of the baseline.

\begin{table}
\centering
  \caption{Comparison of snippet retrieval results in both query directions
  on three types of datasets: (I) fully synthetic, (II) partially real 
  and (III) entirely real.
  Boldfaced rows represent the best performing model per dataset.}
  \label{tab:bl_snip_comparison}
  \scalebox{0.84}{
  \begin{tabular}{lcccc|cccc}
 & \multicolumn{4}{c}{\textbf{Audio-to-Score (A2S)}} & \multicolumn{4}{c}{\textbf{Score-to-Audio (S2A)}} \\ \cmidrule{2-9}
 & \bfseries R@1 & \bfseries R@25 & \bfseries MRR & \bfseries MR &
\bfseries R@1 & \bfseries R@25 & \bfseries MRR & \bfseries MR \\
\midrule
\midrule
\multicolumn{9}{l}{I \ \ \  MSMD (Fully synthetic)} \\
\midrule
BL & 0.54 & 0.91 & 0.653 & 1 & 0.60 & 0.94 & 0.704 & 1 \\
BL+A & \textbf{0.59} & \textbf{0.93} & \textbf{0.699} & \textbf{1} & \textbf{0.61} & \textbf{0.95} & \textbf{0.723} & \textbf{1} \\
BL+S & 0.56 & 0.92 & 0.676 & 1 & 0.61 & 0.94 & 0.717 & 1 \\
BL+A+S & 0.57 & 0.93 & 0.687 & 1 & 0.60 & 0.94 & 0.718 & 1 \\
\midrule
\midrule
\multicolumn{9}{l}{II \ \ RealScores\_Synth (Sheet music scans and synthetic recordings)} \\
\midrule
BL & 0.28 & 0.67 & 0.375 & 7 & 0.36 & 0.77 & 0.467 & 3 \\
BL+A & 0.37 & 0.78 & 0.478 & 3 & 0.43 & 0.82 & 0.537 & 2 \\
BL+S & 0.34 & 0.75 & 0.447 & 4 & 0.43 & 0.84 & 0.544 & 2 \\
BL+A+S & \textbf{0.37} & \textbf{0.79} & \textbf{0.481} & \textbf{3} & \textbf{0.44} & \textbf{0.84 }& \textbf{0.548} & \textbf{2} \\
\midrule
\midrule
\multicolumn{9}{l}{III \ RealScores\_Rec (Sheet music scans and real recordings)} \\
\midrule
BL & 0.10 & 0.36 & 0.156 & 76 & 0.14 & 0.47 & 0.216 & 33 \\
BL+A & 0.13 & 0.44 & 0.200 & 41 & 0.17 & 0.55 & 0.261 & 18 \\
BL+S & 0.12 & 0.42 & 0.192 & 47 & 0.18 & 0.54 & 0.259 & 18 \\
BL+A+S & \textbf{0.15} & \textbf{0.48} & \textbf{0.226} & \textbf{29} & \textbf{0.18} & \textbf{0.54} & \textbf{0.266} & \textbf{18} \\
\end{tabular}}
\end{table}

Table~\ref{tab:bl_snip_comparison} presents the snippet retrieval
results of the four models defined above, evaluated on both search 
directions A2S and S2A. 
In the first section (I) we examine the completely synthetic set 
defined as the MSMD test split.
Then in sections (II) and (III) we consider the partially and 
completely real scenarios, where audio excerpts
are extracted from synthetic and real recordings, respectively,
and sheet music snippets are derived from scans of real scores
in both setups.

We first observe the performance of the current state-of-the-art 
model (\textit{BL}) dropping sharply when moving from 
artificial to real data.
In the fully synthetic set (I) it achieves MRRs of 
0.653 and 0.704 in the directions A2S and S2A, respectively,
correctly retrieving approximately $60\%$ of the snippets 
as the best match in the S2A task.
The MRR drops at least $23\%$ points for either A2S or 
S2A as we move to (II) and at least $48\%$ at (III).
The most extreme drop occurs at (III) in the A2S task: 
only $10\%$ of the score snippets are on rank 1 
(R@1 $= 0.10$).
We additionally note that the retrieval quality of the 
S2A search direction is better than that of A2S for all 
evaluation metrics.

Our proposed models outperform the baseline in all 
scenarios for all evaluation metrics, indicating that
self-supervising pre-training of either modality
is beneficial in the problem we attempt to solve.
We derive the following observations and discussions:
\begin{itemize}
    \item The most significant improvements were observed in 
    configurations with real music data, namely (II) and (III).
    We argue for the modest improvements on (I):
    the synthesized data of MSMD do not exhibit the degradations
    simulated by the augmentations transforms described
    in Sec~\ref{sec:data_augs}, for either scores or recordings.
    Therefore it was not expected that our pre-training strategy
    would considerably benefit retrieval on artificial data.
    \item Pre-training both audio and score encoders (\textit{BL+A+S})
    generated the best retrieval metrics in scenarios with real data, 
    with the largest improvements being observed in (II), where the 
    MRR of the A2S and S2A tasks were increased by roughly $10\%$ and
    $8\%$ points, correspondingly.
    Moreover, it was not observed a substantial compound effect of 
    pre-training both encoders (\textit{BL+A+S}) when comparing to 
    individual encoders (\textit{BL+A} and \textit{BL+S}): the 
    improvements were merely marginal.
    \item In addition to the absolute improvements, the performance drop
    between evaluations on synthesized and real datasets was reduced:
    The MRR gap when moving from (I) to (II) and to (III) reduced by 
    $7.2\%$ and $2.6\%$ points for the A2S direction, respectively; 
    when retrieving the S2A direction these drops accounted for $6.7\%$ 
    and $3.6\%$ points, 
    correspondingly.
    \item The best models also reduced the overall performance gap between 
    retrieval directions A2S and S2A, in all dataset configurations.
\end{itemize}

\begin{figure}[t!]
  \centering
  \includegraphics[width=\linewidth]{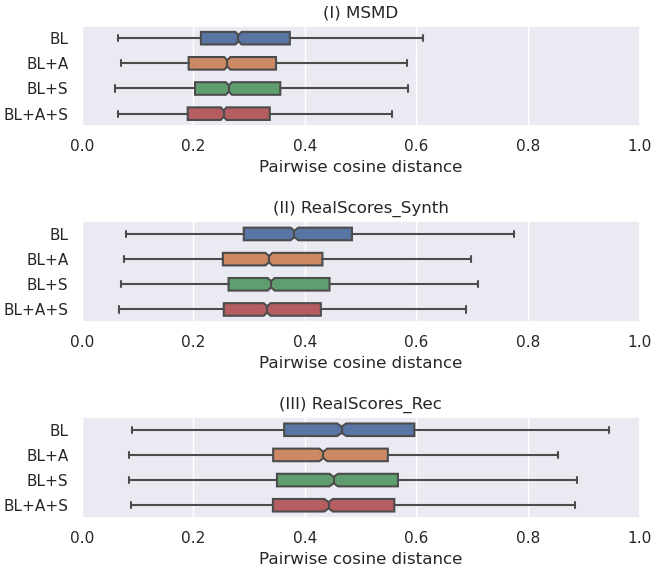}
  \caption{Distributions of pairwise cosine distances between corresponding 
  pairs of audio-sheet music snippets, with 2,000 pairs randomly drawn from 
  each evaluation set. Outliers are not directly visualized in order not
  to distort the plots.
  The vertical lines highlight the medians of the 
  distribution of the baseline model \textit{BL} for each dataset.}
\label{fig:cosdists}
\end{figure}

In an additional experiment, we take a closer look at the shared space 
properties of mapping matching snippets close together.
Figure~\ref{fig:cosdists} depicts the distribution of pairwise cosine 
distances between 2,000 snippet pairs across each test dataset.
When jointly analyzing Table~\ref{tab:bl_snip_comparison}, 
we observe that models which are 
capable of producing smaller distances between matching fragments
generate better snippet retrieval quality.
Moreover, we see that distances between snippet pairs from real 
data are on average 
mapped farther to each other than those from synthesized music data.

In all experimental scenarios, our pre-trained models were able to 
project corresponding snippets in the embedding space closer together,
in comparison with the state-of-the-art method.
From this we can point out the potential of self-supervised pre-training 
as a key component towards more powerful joint embedding spaces.

\subsection{Cross-modal Piece Identification Experiments}
\label{subsec:exp_2}

In this set of experiments, we aggregate snippet embeddings generated
by our models to perform cross-modal piece identification:
given a full recording as a query, retrieve the corresponding score 
from a collection; or given a printed score, find an appropriate 
music recording within a database.
We evaluate this task as a proof-of-concept, to validate our proposed 
methods in a higher-level realistic retrieval scenario.
As underlined in Section~\ref{sec:intro}, piece identification is a key
component of many audio--score retrieval systems, so we believe this
evaluation can give us insights towards more robust systems.

The piece identification is done as in~\cite{DorferHAFW18_MSMD_TISMIR},
with an approach that we will denote as \textit{vote-based}: a matching 
procedure purely based on snippet retrieval and indexing.
Let $ \mathcal{D} $ be a collection of $L$ documents and $Q$ be a document 
query in the target and search modalities, respectively.
Each document $ D_i \in \mathcal{D} $ is sequentially cut into snippets, 
which are embedded via their respective pathway network of 
Figure~\ref{fig:bl_schema}, generating a set of embeddings 
$ \{ d^i_1, d^i_2, ..., d^i_{M_i} \} $, where each embedding $d^i_j$ is 
indexed to its originating document $D_i$.
We define hop sizes of 50 pixels and 10 frames (roughly 0.5 sec) for 
consecutive sheet music and audio snippets.

The document query is segmented into 100 equally-spaced excerpts, which are 
passed through the counterpart pathway of the model, producing a
set of query embeddings $ \{ q_1, q_2, ..., q_{100} \} $.
Then snippet retrieval is carried out as in Section~\ref{subsec:exp_1} 
for each query embedding $q_j$, with the difference that now the top 
25 nearest neighbors are retrieved per query embedding among all 
embeddings from the collection $ \mathcal{D} $.
Each nearest neighbor votes for its originating document, and a 
vote-based ranked list is created by aggregating all nearest neighbors 
from all 100 query embeddings.
The document achieving the highest count among all 2,500 votes is selected 
as the best match.

\begin{table*}[t!]
\centering
  \caption{Comparison of audio--sheet music piece identification results 
  in both query directions on three types of datasets: (I) fully synthetic, (II) partially real 
  and (III) entirely real.
  Boldfaced rows represent the best performing model per dataset.}
  \label{tab:piece_ret_comparison}
  \scalebox{1}{
 \begin{tabular}{llcccc|cccc}
 & & \multicolumn{4}{c}{\textbf{Audio-to-Score (A2S)}} & \multicolumn{4}{c}{\textbf{Score-to-Audio (S2A)}} \\ \cmidrule{3-10}
 & \bfseries \# & \bfseries R@1 & \bfseries R@10 & \bfseries $>$R@10 & \bfseries MRR &
\bfseries R@1 & \bfseries R@10 & \bfseries $>$R@10 & \bfseries MRR \\
\midrule
\midrule
\multicolumn{10}{l}{I \ \ \  MSMD (Fully synthetic)} \\
\midrule
BL & 100 & 0.76 (76) & 0.98 (98) & 0.02 (2) & 0.846  & 0.87 (87) & 1.00 (100) & 0.00 (0) & 0.927  \\
BL+A & 100 & 0.85 (85) & 0.99 (99) & 0.01 (1) & 0.910  & 0.81 (81) & 1.00 (100) & 0.00 (0) & 0.896  \\
BL+S & 100 & 0.84 (84) & 1.00 (100) & 0.00 (0) & 0.898  & 0.87 (87) & 1.00 (100) & 0.00 (0) & 0.928  \\
BL+A+S & 100 & \textbf{0.87 (87)} & \textbf{1.00 (100)} & \textbf{0.00 (0)} & \textbf{0.918}  & \textbf{0.93 (93)} & \textbf{1.00 (100)} & \textbf{0.00 (0)} & \textbf{0.961}  \\
\midrule
\midrule
\multicolumn{10}{l}{II \ \ RealScores\_Synth (Sheet music scans and synthetic recordings)} \\
\midrule
BL & 314 & 0.49 (154) & 0.84 (265) & 0.16 (49) & 0.609  & 0.65 (203) & 0.90 (282) & 0.10 (32) & 0.734  \\
BL+A & 314 & 0.71 (223) & 0.94 (294) & 0.06 (20) & 0.792  & 0.82 (256) & 0.98 (307) & 0.02 (7) & 0.874  \\
BL+S & 314 & 0.70 (219) & 0.93 (291) & 0.07 (23) & 0.781  & 0.82 (256) & 0.97 (306) & 0.03 (8) & 0.871  \\
BL+A+S & 314 & \textbf{0.80 (250)} & \textbf{0.96 (302)} & \textbf{0.04 (12)} & \textbf{0.857}  & \textbf{0.88 (277)} & \textbf{0.98 (308)} & \textbf{0.02 (6)} & \textbf{0.919}  \\
\midrule
\midrule
\multicolumn{10}{l}{III \ RealScores\_Rec (Sheet music scans and real recordings)} \\
\midrule
BL & 198 & 0.11 (22) & 0.57 (113) & 0.43 (85) & 0.256  & 0.48 (95) & 0.79 (156) & 0.21 (42) & 0.587  \\
BL+A & 198 & 0.21 (42) & 0.69 (136) & 0.31 (62) & 0.361  & 0.62 (122) & 0.87 (173) & 0.13 (25) & 0.714  \\
BL+S & 198 & 0.22 (44) & 0.69 (137) & 0.31 (61) & 0.375  & 0.63 (125) & 0.88 (175) & 0.12 (23) & 0.721  \\
BL+A+S & 198 & \textbf{0.39 (78)} & \textbf{0.81 (161)} & \textbf{0.19 (37)} & \textbf{0.535}  & \textbf{0.72 (143)} & \textbf{0.94 (187)} & \textbf{0.06 (11)} & \textbf{0.795}  \\
\end{tabular}}
\end{table*}

In our piece identification experiments we evaluate on pieces of the
same datasets as in Section~\ref{subsec:exp_1}.
(I) The MSMD test split has 100 pairs of both synthesized scores and their
respective recordings; (II) has 314 pieces with their corresponding 
scanned sheet music and synthesized recordings; and (III) has 198 pairs 
of scanned sheet music and real recordings.

The cross-modal piece identification results are summarized in
Table~\ref{tab:piece_ret_comparison}.
We evaluate the same scenarios and models as for the two-way snippet
retrieval task, in both search directions A2S and S2A.
Moreover we include in the table (between parentheses) the actual 
number of pieces retrieved for each recall value.

The piece identification results exhibit a similar trend as in 
the previous experiments on snippet retrieval.
The performance of the baseline model \textit{BL} also declines
abruptly as more real scenarios are evaluated.
The mean reciprocal rank drops around $59\%$ and $34\%$ points 
when traversing from (I) to the most realistic case (III), for the
retrieval directions A2S and S2A, correspondingly.
The worst case happens at (III) for the A2S direction, when only 
approximately $11\%$ of the scores (22 items among 198) are 
correctly retrieved as the best match.

We derive the following discussions and observations concerning the 
performance of our proposed methods:
\begin{itemize}
    \item In configurations with real data, our methods outperformed 
    the baseline \textit{BL} in all evaluation metrics by a significant
    margin, with \textit{BL+A+S} being the best model among them.
    For example, in the fully real scenario (III)
    the MRR of \textit{BL+A+S} in the A2S direction 
    increased from 0.256 to 0.535, which indicates a performance
    jump of more than $100\%$ of the former value; in the S2A 
    direction, now only $6\%$ of the recordings are not correctly 
    retrieved among the best ten matches.
    \item The compound effect of pre-training both encoders (\textit{BL+A+S}) when comparing to individual encoders (\textit{BL+A} and \textit{BL+S}) was 
    stronger than in the two-way snippet retrieval. In the 
    (III)--(\textit{BL+A+S})--(A2S) configuration the MRR improvement accounted
    for more than the sum of the individual improvements observed for models
    \textit{BL+A} and \textit{BL+S}.
    \item In addition to the dataset-wise improvements, the performance gaps 
    between synthesized and real datasets, and between A2S and S2A directions,
    were significantly reduced.
    \item Overall the boost in retrieval quality that our proposed models produced
    is significantly higher for cross-modal piece identification than for 
    snippet retrieval.
    This indicates that a moderate performance boost in short fragment-level 
    music retrieval tasks has great potential to escalate to greater improvements
    in higher-level retrieval problems if a proper post-processing method 
    aggregating those fragments is employed.
\end{itemize}

To get a better understanding of the matching quality of our models on 
piece identification scenarios, we discuss on the 
\textit{separation indicator}, introduced
in~\cite{ZalkowBM19_SalienceRetrieval_ICASSP}.
This factor measures how distinct the relevant document is among the 
other items during the retrieval process.
Given the vote-based ranked list created during the identification 
procedure of query $Q$, its counterpart document is retrieved at 
rank $r$.
Defining $\delta_{D_i}$ as the number of votes the document ranked at 
$i\mathrm{-th}$ position received, the separation indicator 
$\rho \in \mathbb{R}_{+}$ is defined as:
\begin{equation}
    \rho =
    \begin{cases}
      \delta_{D_2} / \delta_{D_1} & \text{if rank $r=1$,}\\
      \delta_{D_1} / \delta_{D_r} & \text{otherwise.}
    \end{cases}
\end{equation}
In this metric, indicators below $1$ point out to a correct match, with 
lower values indicating better matching quality. A $\rho > 1$ implies
a wrong detection; the bigger its value, the lesser is number of votes 
received by the correct document in comparison with the top match.

\begin{figure*}[t!]
  \centering
  \includegraphics[width=\textwidth]{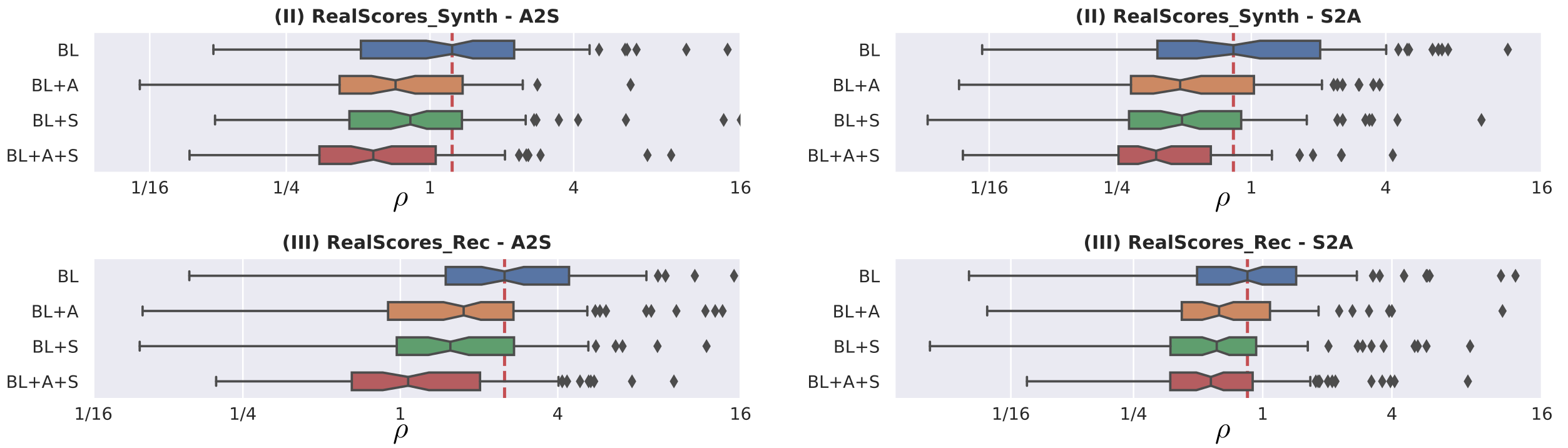}
  \caption{Distributions of the separation indicators produced for 
  cross-modal piece identification on datasets (II) and (III), reducing
  the number of pieces to 100 per set.
  The horizontal axes are displayed in logarithmic scale.}
\label{fig:mqual}
\end{figure*}

Figure~\ref{fig:mqual} visualizes the distribution of the separation 
indicators obtained when performing cross-modal piece identification
on the datasets with real music data.
In this experiment we reduce the number of documents of each dataset 
to 100 pairs of audio recordings and scanned scores.

%
A joint analysis with Table~\ref{tab:bl_snip_comparison} reveals
that overall the models with better piece identification results 
also exhibit a better matching quality statistics.
Noteworthy is the poor matching quality of the (III)-A2S setup, 
the most realistic case in the audio--score search direction: the 
distributions of all models are strongly concentrated above $\rho = 1$.
Our proposed methods generated overall smaller separation indicators for
all audio--sheet music identification setups, indicating that 
self-supervised learning is a promising orientation for reliable 
audio--score retrieval systems.


\section{Conclusion}
\label{sec:conclusion}

In this work we designed a learning framework to alleviate 
labeled data scarcity in training networks to solve audio--score 
retrieval tasks.
We proposed multi-modal self-supervised contrastive learning 
of short excerpts of sheet music images and audio recordings as
a first pre-processing step.
In this framework, the network responsible for encoding each 
modality can be independently pre-trained and enabled for
fine-tuning, having the potential to adapt to different tasks
that require different fine-tuning configurations.
For that we define a pipeline of augmentation transforms specifically for
audio and sheet music snippets, and employ publicly available 
music data to pre-train our networks.

Experiments on two-way snippet retrieval and subsequently 
on cross-modal piece identification evaluating diverse datasets 
showed that our proposed framework outperforms current state-of-the 
arts methods, specially in scenarios composed 
partially or entirely of real music data. 
Moreover, the self-supervised approach helped reducing the performance
gap between synthetic and real data, which is one of the main 
challenges of audio--score retrieval problems.

Given the improved retrieval performance in realist configurations,
in addition to larges amounts of publicly available music data 
that are available with easy access, we believe this is a promising 
research direction for the design of robust multi-modal music 
search and retrieval systems.

\begin{acks}
This work is supported by the European Research Council
(ERC) under the EU’s Horizon 2020 research and innovation
programme, grant agreement No. 101019375 (“Whither
Music?”), and the Federal State of Upper Austria (LIT AI
Lab). 
\end{acks}

\bibliographystyle{ACM-Reference-Format}
\bibliography{manuscript}


\begin{thebibliography}{42}


\ifx \showCODEN    \undefined \def \showCODEN     #1{\unskip}     \fi
\ifx \showDOI      \undefined \def \showDOI       #1{#1}\fi
\ifx \showISBNx    \undefined \def \showISBNx     #1{\unskip}     \fi
\ifx \showISBNxiii \undefined \def \showISBNxiii  #1{\unskip}     \fi
\ifx \showISSN     \undefined \def \showISSN      #1{\unskip}     \fi
\ifx \showLCCN     \undefined \def \showLCCN      #1{\unskip}     \fi
\ifx \shownote     \undefined \def \shownote      #1{#1}          \fi
\ifx \showarticletitle \undefined \def \showarticletitle #1{#1}   \fi
\ifx \showURL      \undefined \def \showURL       {\relax}        \fi
\providecommand\bibfield[2]{#2}
\providecommand\bibinfo[2]{#2}
\providecommand\natexlab[1]{#1}
\providecommand\showeprint[2][]{arXiv:#2}

\bibitem[Agarap(2018)]%
        {Agarap18_ReLU_CVPR}
\bibfield{author}{\bibinfo{person}{Abien~Fred Agarap}.}
  \bibinfo{year}{2018}\natexlab{}.
\newblock \bibinfo{title}{Deep Learning using Rectified Linear Units ({ReLU})}.
\newblock
\newblock
\urldef\tempurl%
\url{https://doi.org/10.48550/ARXIV.1803.08375}
\showDOI{\tempurl}


\bibitem[Agrawal et~al\mbox{.}(2022)]%
        {AgrawalWD21_CNNAttentionAlignment_IEEE-SPL}
\bibfield{author}{\bibinfo{person}{Ruchit Agrawal}, \bibinfo{person}{Daniel
  Wolff}, {and} \bibinfo{person}{Simon Dixon}.}
  \bibinfo{year}{2022}\natexlab{}.
\newblock \showarticletitle{A Convolutional-Attentional Neural Framework for
  Structure-Aware Performance-Score Synchronization}.
\newblock \bibinfo{journal}{\emph{IEEE Signal Processing Letters}}
  \bibinfo{volume}{29} (\bibinfo{year}{2022}), \bibinfo{pages}{344--348}.
\newblock


\bibitem[Arzt et~al\mbox{.}(2014)]%
        {ArztBFFGLW14_PianoCompanion_PAIS}
\bibfield{author}{\bibinfo{person}{Andreas Arzt}, \bibinfo{person}{Sebastian
  B{\"o}ck}, \bibinfo{person}{Sebastian Flossmann}, \bibinfo{person}{Harald
  Frostel}, \bibinfo{person}{Martin Gasser}, \bibinfo{person}{Cynthia~C.S.
  Liem}, {and} \bibinfo{person}{Gerhard Widmer}.}
  \bibinfo{year}{2014}\natexlab{}.
\newblock \showarticletitle{The Piano Music Companion}. In
  \bibinfo{booktitle}{\emph{Proceedings of the Conference on Prestigious
  Applications of Intelligent Systems (PAIS)}}. \bibinfo{address}{Prague,
  Czechia}.
\newblock


\bibitem[Arzt et~al\mbox{.}(2012)]%
        {ArztBW12_SymbolicFingerprint_ISMIR}
\bibfield{author}{\bibinfo{person}{Andreas Arzt}, \bibinfo{person}{Sebastian
  B{\"o}ck}, {and} \bibinfo{person}{Gerhard Widmer}.}
  \bibinfo{year}{2012}\natexlab{}.
\newblock \showarticletitle{Fast Identification of Piece and Score Position via
  Symbolic Fingerprinting}. In \bibinfo{booktitle}{\emph{Proceedings of the
  International Society for Music Information Retrieval Conference ({ISMIR})}}.
  \bibinfo{address}{Porto, Portugal}, \bibinfo{pages}{433--438}.
\newblock


\bibitem[Arzt et~al\mbox{.}(2008)]%
        {ArztWD08_PageTurning_ECAI}
\bibfield{author}{\bibinfo{person}{Andreas Arzt}, \bibinfo{person}{Gerhard
  Widmer}, {and} \bibinfo{person}{Simon Dixon}.}
  \bibinfo{year}{2008}\natexlab{}.
\newblock \showarticletitle{Automatic Page Turning for Musicians via Real-Time
  Machine Listening}. In \bibinfo{booktitle}{\emph{In Proceedings of the 18th
  European Conference on Artificial Intelligence (ECAI)}}.
  \bibinfo{address}{Patras, Greece}, \bibinfo{pages}{241--245}.
\newblock


\bibitem[Balke et~al\mbox{.}(2016)]%
        {BalkeALM16_BarlowRetrieval_ICASSP}
\bibfield{author}{\bibinfo{person}{Stefan Balke}, \bibinfo{person}{Vlora
  Arifi-M{\"u}ller}, \bibinfo{person}{Lukas Lamprecht}, {and}
  \bibinfo{person}{Meinard M{\"u}ller}.} \bibinfo{year}{2016}\natexlab{}.
\newblock \showarticletitle{Retrieving Audio Recordings Using Musical Themes}.
  In \bibinfo{booktitle}{\emph{Proceedings of the {IEEE} International
  Conference on Acoustics, Speech, and Signal Processing ({ICASSP})}}.
  \bibinfo{address}{Shanghai, China}, \bibinfo{pages}{281--285}.
\newblock


\bibitem[Balke et~al\mbox{.}(2019)]%
        {BalkeDCAW19_ASR_TempoInv_ISMIR}
\bibfield{author}{\bibinfo{person}{Stefan Balke}, \bibinfo{person}{Matthias
  Dorfer}, \bibinfo{person}{Luis Carvalho}, \bibinfo{person}{Andreas Arzt},
  {and} \bibinfo{person}{Gerhard Widmer}.} \bibinfo{year}{2019}\natexlab{}.
\newblock \showarticletitle{Learning Soft-Attention Models for Tempo-invariant
  Audio-Sheet Music Retrieval}. In \bibinfo{booktitle}{\emph{Proceedings of the
  International Society for Music Information Retrieval Conference (ISMIR)}}.
  \bibinfo{address}{Delft, Netherlands}, \bibinfo{pages}{216--222}.
\newblock


\bibitem[B{\"{o}}ck and Schedl(2012)]%
        {BoeckS12_TranscriptionRecurrentNetwork_ICASSP}
\bibfield{author}{\bibinfo{person}{Sebastian B{\"{o}}ck} {and}
  \bibinfo{person}{Markus Schedl}.} \bibinfo{year}{2012}\natexlab{}.
\newblock \showarticletitle{Polyphonic piano note transcription with recurrent
  neural networks}. In \bibinfo{booktitle}{\emph{{IEEE} International
  Conference on Acoustics, Speech and Signal Processing ({ICASSP})}}.
  \bibinfo{pages}{121--124}.
\newblock


\bibitem[Calvo-Zaragoza et~al\mbox{.}(2021)]%
        {Calvo-ZaragozaHP21_OMRReview_ACM}
\bibfield{author}{\bibinfo{person}{Jorge Calvo-Zaragoza},
  \bibinfo{person}{Jan~Haji\v{c} Jr.}, {and} \bibinfo{person}{Alexander
  Pacha}.} \bibinfo{year}{2021}\natexlab{}.
\newblock \showarticletitle{Understanding Optical Music Recognition}.
\newblock \bibinfo{journal}{\emph{Comput. Surveys}} \bibinfo{volume}{53},
  \bibinfo{number}{4} (\bibinfo{year}{2021}).
\newblock


\bibitem[Cancino-Chac{\'o}n et~al\mbox{.}(2017)]%
        {CancinoChaconGWG_Magaloff_ML}
\bibfield{author}{\bibinfo{person}{Carlos~Eduardo Cancino-Chac{\'o}n},
  \bibinfo{person}{Thassilo Gadermaier}, \bibinfo{person}{Gerhard Widmer},
  {and} \bibinfo{person}{Maarten Grachten}.} \bibinfo{year}{2017}\natexlab{}.
\newblock \showarticletitle{An evaluation of linear and non-linear models of
  expressive dynamics in classical piano and symphonic music}.
\newblock \bibinfo{journal}{\emph{Machine Learning}} \bibinfo{volume}{106},
  \bibinfo{number}{6} (\bibinfo{year}{2017}), \bibinfo{pages}{887--909}.
\newblock


\bibitem[Chen et~al\mbox{.}(2020)]%
        {ChenKNH20_SimCLR_ICML}
\bibfield{author}{\bibinfo{person}{Ting Chen}, \bibinfo{person}{Simon
  Kornblith}, \bibinfo{person}{Mohammad Norouzi}, {and}
  \bibinfo{person}{Geoffrey Hinton}.} \bibinfo{year}{2020}\natexlab{}.
\newblock \showarticletitle{A Simple Framework for Contrastive Learning of
  Visual Representations}. In \bibinfo{booktitle}{\emph{Proceedings of the 37th
  International Conference on Machine Learning ({ICML})}}.
\newblock


\bibitem[Chopra et~al\mbox{.}(2005)]%
        {ChopraHL05_SiameseNNs_CVPR}
\bibfield{author}{\bibinfo{person}{Sumit Chopra}, \bibinfo{person}{Raia
  Hadsell}, {and} \bibinfo{person}{Yann LeCun}.}
  \bibinfo{year}{2005}\natexlab{}.
\newblock \showarticletitle{Learning a similarity metric discriminatively, with
  application to face verification}. In \bibinfo{booktitle}{\emph{Proceedings
  of the {IEEE} Conference on Computer Vision and Pattern Recognition
  ({CVPR})}}, Vol.~\bibinfo{volume}{1}. \bibinfo{pages}{539--546}.
\newblock


\bibitem[Clevert et~al\mbox{.}(2016)]%
        {ClevertUH16_ELU_ICLR}
\bibfield{author}{\bibinfo{person}{Djork{-}Arn{\'{e}} Clevert},
  \bibinfo{person}{Thomas Unterthiner}, {and} \bibinfo{person}{Sepp
  Hochreiter}.} \bibinfo{year}{2016}\natexlab{}.
\newblock \showarticletitle{Fast and Accurate Deep Network Learning by
  Exponential Linear Units ({ELUs})}. In
  \bibinfo{booktitle}{\emph{International Conferen1ce on Learning
  Representations, (ICLR)}}.
\newblock


\bibitem[Dorfer et~al\mbox{.}(2017)]%
        {DorferAW17_ScoreIdentification_ISMIR}
\bibfield{author}{\bibinfo{person}{Matthias Dorfer}, \bibinfo{person}{Andreas
  Arzt}, {and} \bibinfo{person}{Gerhard Widmer}.}
  \bibinfo{year}{2017}\natexlab{}.
\newblock \showarticletitle{Learning Audio-Sheet Music Correspondences for
  Score Identification and Offline Alignment}. In
  \bibinfo{booktitle}{\emph{Proceedings of the International Society for Music
  Information Retrieval Conference ({ISMIR})}}. \bibinfo{address}{Suzhou,
  China}, \bibinfo{pages}{115--122}.
\newblock


\bibitem[Dorfer et~al\mbox{.}(2018a)]%
        {DorferHAFW18_MSMD_TISMIR}
\bibfield{author}{\bibinfo{person}{Matthias Dorfer}, \bibinfo{person}{Jan
  {Haji{\v{c}} jr.}}, \bibinfo{person}{Andreas Arzt}, \bibinfo{person}{Harald
  Frostel}, {and} \bibinfo{person}{Gerhard Widmer}.}
  \bibinfo{year}{2018}\natexlab{a}.
\newblock \showarticletitle{Learning Audio--Sheet Music Correspondences for
  Cross-Modal Retrieval and Piece Identification}.
\newblock \bibinfo{journal}{\emph{Transactions of the International Society for
  Music Information Retrieval}} \bibinfo{volume}{1}, \bibinfo{number}{1}
  (\bibinfo{year}{2018}).
\newblock


\bibitem[Dorfer et~al\mbox{.}(2018b)]%
        {DorferSVKW18_CCALayer_IJMIR}
\bibfield{author}{\bibinfo{person}{Matthias Dorfer}, \bibinfo{person}{Jan
  Schl{\"u}ter}, \bibinfo{person}{Andreu Vall}, \bibinfo{person}{Filip
  Korzeniowski}, {and} \bibinfo{person}{Gerhard Widmer}.}
  \bibinfo{year}{2018}\natexlab{b}.
\newblock \showarticletitle{End-to-end cross-modality retrieval with {CCA}
  projections and pairwise ranking loss}.
\newblock \bibinfo{journal}{\emph{International Journal of Multimedia
  Information Retrieval}} \bibinfo{volume}{7}, \bibinfo{number}{2}
  (\bibinfo{date}{01 6} \bibinfo{year}{2018}), \bibinfo{pages}{117--128}.
\newblock
\showISSN{2192-662X}


\bibitem[Dosovitskiy et~al\mbox{.}(2014)]%
        {DosovitskiySRB14_InstanceDiscrimination_NIPS}
\bibfield{author}{\bibinfo{person}{Alexey Dosovitskiy},
  \bibinfo{person}{Jost~Tobias Springenberg}, \bibinfo{person}{Martin
  Riedmiller}, {and} \bibinfo{person}{Thomas Brox}.}
  \bibinfo{year}{2014}\natexlab{}.
\newblock \showarticletitle{Discriminative Unsupervised Feature Learning with
  Convolutional Neural Networks}. In \bibinfo{booktitle}{\emph{Advances in
  Neural Information Processing Systems}}, Vol.~\bibinfo{volume}{27}.
\newblock


\bibitem[Fremerey et~al\mbox{.}(2009)]%
        {FremereyCME09_SheetMusicID_ISMIR}
\bibfield{author}{\bibinfo{person}{Christian Fremerey},
  \bibinfo{person}{Michael Clausen}, \bibinfo{person}{Sebastian Ewert}, {and}
  \bibinfo{person}{Meinard M{\"u}ller}.} \bibinfo{year}{2009}\natexlab{}.
\newblock \showarticletitle{Sheet Music-Audio Identification}. In
  \bibinfo{booktitle}{\emph{Proceedings of the International Conference on
  Music Information Retrieval ({ISMIR})}}. \bibinfo{address}{Kobe, Japan},
  \bibinfo{pages}{645--650}.
\newblock


\bibitem[Hawthorne et~al\mbox{.}(2019)]%
        {HawthorneSR+_MAESTRO_ICLR}
\bibfield{author}{\bibinfo{person}{Curtis Hawthorne}, \bibinfo{person}{Andriy
  Stasyuk}, \bibinfo{person}{Adam Roberts}, \bibinfo{person}{Ian Simon},
  \bibinfo{person}{Cheng-Zhi~Anna Huang}, \bibinfo{person}{Sander Dieleman},
  \bibinfo{person}{Erich Elsen}, \bibinfo{person}{Jesse Engel}, {and}
  \bibinfo{person}{Douglas Eck}.} \bibinfo{year}{2019}\natexlab{}.
\newblock \showarticletitle{Enabling Factorized Piano Music Modeling and
  Generation with the {MAESTRO} Dataset}. In
  \bibinfo{booktitle}{\emph{International Conference on Learning
  Representations}}.
\newblock


\bibitem[He et~al\mbox{.}(2015)]%
        {HeZRS15_HeInit_ICCV}
\bibfield{author}{\bibinfo{person}{Kaiming He}, \bibinfo{person}{Xiangyu
  Zhang}, \bibinfo{person}{Shaoqing Ren}, {and} \bibinfo{person}{Jian Sun}.}
  \bibinfo{year}{2015}\natexlab{}.
\newblock \showarticletitle{Delving Deep into Rectifiers: Surpassing
  Human-Level Performance on ImageNet Classification}. In
  \bibinfo{booktitle}{\emph{IEEE International Conference on Computer Vision
  ({ICCV})}}. \bibinfo{pages}{1026--1034}.
\newblock


\bibitem[Henkel and Widmer(2021)]%
        {HenkelW21_RTScoreFollowing_Frontiers}
\bibfield{author}{\bibinfo{person}{Florian Henkel} {and}
  \bibinfo{person}{Gerhard Widmer}.} \bibinfo{year}{2021}\natexlab{}.
\newblock \showarticletitle{Real-Time Music Following in Score Sheet Images via
  Multi-Resolution Prediction}.
\newblock \bibinfo{journal}{\emph{Frontiers in Computer Science}}
  \bibinfo{volume}{3} (\bibinfo{year}{2021}).
\newblock


\bibitem[Ioffe and Szegedy(2015)]%
        {IoffeS15_BatchNorm_ICML}
\bibfield{author}{\bibinfo{person}{Sergey Ioffe} {and}
  \bibinfo{person}{Christian Szegedy}.} \bibinfo{year}{2015}\natexlab{}.
\newblock \showarticletitle{Batch Normalization: Accelerating Deep Network
  Training by Reducing Internal Covariate Shift}. In
  \bibinfo{booktitle}{\emph{Proceedings of the 32nd International Conference on
  International Conference on Machine Learning (ICML)}}.
  \bibinfo{address}{Lille, France}, \bibinfo{pages}{448--456}.
\newblock


\bibitem[Izmirli and Sharma(2012)]%
        {IzmirliS12_PrintedMusicAudio_ISMIR}
\bibfield{author}{\bibinfo{person}{{\"{O}}zg{\"{u}}r Izmirli} {and}
  \bibinfo{person}{Gyanendra Sharma}.} \bibinfo{year}{2012}\natexlab{}.
\newblock \showarticletitle{Bridging Printed Music and Audio Through Alignment
  Using a Mid-level Score Representation}. In
  \bibinfo{booktitle}{\emph{Proceedings of the International Society for Music
  Information Retrieval Conference ({ISMIR})}}. \bibinfo{address}{Porto,
  Portugal}, \bibinfo{pages}{61--66}.
\newblock
\urldef\tempurl%
\url{http://ismir2012.ismir.net/event/papers/061-ismir-2012.pdf}
\showURL{%
\tempurl}


\bibitem[Kingma and Ba(2015)]%
        {KingmaB15_Adam_ICLR}
\bibfield{author}{\bibinfo{person}{Diederik~P. Kingma} {and}
  \bibinfo{person}{Jimmy Ba}.} \bibinfo{year}{2015}\natexlab{}.
\newblock \showarticletitle{Adam: A Method for Stochastic Optimization}. In
  \bibinfo{booktitle}{\emph{International Conference on Learning
  Representations ({ICLR})}}.
\newblock


\bibitem[Kiros et~al\mbox{.}(2014)]%
        {KirosSZ14_VisualSemanticEmbeddings_arxiv}
\bibfield{author}{\bibinfo{person}{Ryan Kiros}, \bibinfo{person}{Ruslan
  Salakhutdinov}, {and} \bibinfo{person}{Richard~S. Zemel}.}
  \bibinfo{year}{2014}\natexlab{}.
\newblock \showarticletitle{Unifying Visual-Semantic Embeddings with Multimodal
  Neural Language Models}.
\newblock \bibinfo{journal}{\emph{arXiv preprint (arXiv:1411.2539)}}
  (\bibinfo{year}{2014}).
\newblock


\bibitem[Krizhevsky et~al\mbox{.}(2012)]%
        {KrizhevskySH12_ImageNet_NIPS}
\bibfield{author}{\bibinfo{person}{Alex Krizhevsky}, \bibinfo{person}{Ilya
  Sutskever}, {and} \bibinfo{person}{Geoffrey~E. Hinton}.}
  \bibinfo{year}{2012}\natexlab{}.
\newblock \showarticletitle{ImageNet Classification with Deep Convolutional
  Neural Networks}. In \bibinfo{booktitle}{\emph{Advances in Neural Information
  Processing Systems}}. \bibinfo{pages}{1097--1105}.
\newblock


\bibitem[L{\'o}pez-Guti{\'e}rrez et~al\mbox{.}(2021)]%
        {LopezVCC21_OMRAug_ICDAR}
\bibfield{author}{\bibinfo{person}{Juan~C. L{\'o}pez-Guti{\'e}rrez},
  \bibinfo{person}{Jose~J. Valero-Mas}, \bibinfo{person}{Francisco~J.
  Castellanos}, {and} \bibinfo{person}{Jorge Calvo-Zaragoza}.}
  \bibinfo{year}{2021}\natexlab{}.
\newblock \showarticletitle{Data Augmentation for End-to-End Optical Music
  Recognition}. In \bibinfo{booktitle}{\emph{Proceedings of the 14th IAPR
  International Workshop on Graphics Recognition ({GREC})}}.
  \bibinfo{publisher}{Springer}, \bibinfo{pages}{59--73}.
\newblock


\bibitem[M{\"u}ller et~al\mbox{.}(2019)]%
        {MuellerABDW19_MusicRetrieval_IEEE-SPM}
\bibfield{author}{\bibinfo{person}{Meinard M{\"u}ller},
  \bibinfo{person}{Andreas Arzt}, \bibinfo{person}{Stefan Balke},
  \bibinfo{person}{Matthias Dorfer}, {and} \bibinfo{person}{Gerhard Widmer}.}
  \bibinfo{year}{2019}\natexlab{}.
\newblock \showarticletitle{Cross-Modal Music Retrieval and Applications: An
  Overview of Key Methodologies}.
\newblock \bibinfo{journal}{\emph{{IEEE} Signal Processing Magazine}}
  \bibinfo{volume}{36}, \bibinfo{number}{1} (\bibinfo{year}{2019}),
  \bibinfo{pages}{52--62}.
\newblock


\bibitem[Park et~al\mbox{.}(2019)]%
        {ParkCZCZCL19_SpecAugment_INTERSPEECH}
\bibfield{author}{\bibinfo{person}{Daniel~S. Park}, \bibinfo{person}{William
  Chan}, \bibinfo{person}{Yu Zhang}, \bibinfo{person}{Chung{-}Cheng Chiu},
  \bibinfo{person}{Barret Zoph}, \bibinfo{person}{Ekin~D. Cubuk}, {and}
  \bibinfo{person}{Quoc~V. Le}.} \bibinfo{year}{2019}\natexlab{}.
\newblock \showarticletitle{SpecAugment: {A} Simple Data Augmentation Method
  for Automatic Speech Recognition}. In \bibinfo{booktitle}{\emph{Proceedings
  of the Annual Conference of the International Speech Communication
  Association ({INTERSPEECH})}}. \bibinfo{pages}{2613--2617}.
\newblock


\bibitem[Perlin(2002)]%
        {Perlin02_PerlinNoise_ACMTOG}
\bibfield{author}{\bibinfo{person}{Ken Perlin}.}
  \bibinfo{year}{2002}\natexlab{}.
\newblock \showarticletitle{Improving Noise}.
\newblock \bibinfo{journal}{\emph{ACM Transactions on Graphics ({TOG})}}
  \bibinfo{volume}{21}, \bibinfo{number}{3} (\bibinfo{year}{2002}),
  \bibinfo{pages}{681--682}.
\newblock


\bibitem[Salamon and Bello(2017)]%
        {SalamonB17_CNNDataAugForEnvSoundClass_IEEE-LettersSP}
\bibfield{author}{\bibinfo{person}{Justin Salamon} {and}
  \bibinfo{person}{Juan~Pablo Bello}.} \bibinfo{year}{2017}\natexlab{}.
\newblock \showarticletitle{Deep Convolutional Neural Networks and Data
  Augmentation for Environmental Sound Classification}.
\newblock \bibinfo{journal}{\emph{{IEEE} Signal Processing Letters}}
  \bibinfo{volume}{24}, \bibinfo{number}{3} (\bibinfo{year}{2017}),
  \bibinfo{pages}{279--283}.
\newblock


\bibitem[Schl{\"{u}}ter and Grill(2015)]%
        {SchlueterG15_DataAug_ISMIR}
\bibfield{author}{\bibinfo{person}{Jan Schl{\"{u}}ter} {and}
  \bibinfo{person}{Thomas Grill}.} \bibinfo{year}{2015}\natexlab{}.
\newblock \showarticletitle{Exploring Data Augmentation for Improved Singing
  Voice Detection with Neural Networks}. In
  \bibinfo{booktitle}{\emph{Proceedings of the International Society for Music
  Information Retrieval Conference ({ISMIR})}}. \bibinfo{pages}{121--126}.
\newblock


\bibitem[Shorten and Khoshgoftaar(2019)]%
        {ShortenK19_DataAugSurvey_JBD}
\bibfield{author}{\bibinfo{person}{Connor Shorten} {and}
  \bibinfo{person}{Taghi~M. Khoshgoftaar}.} \bibinfo{year}{2019}\natexlab{}.
\newblock \showarticletitle{A survey on Image Data Augmentation for Deep
  Learning}.
\newblock \bibinfo{journal}{\emph{Journal of Big Data}} \bibinfo{volume}{6},
  \bibinfo{number}{60} (\bibinfo{year}{2019}).
\newblock
Issue 1.


\bibitem[Sigtia et~al\mbox{.}(2016)]%
        {SigtiaBD16_DNNPolyPianoTrans_TASLP}
\bibfield{author}{\bibinfo{person}{Siddharth Sigtia},
  \bibinfo{person}{Emmanouil Benetos}, {and} \bibinfo{person}{Simon Dixon}.}
  \bibinfo{year}{2016}\natexlab{}.
\newblock \showarticletitle{An End-to-End Neural Network for Polyphonic Piano
  Music Transcription}.
\newblock \bibinfo{journal}{\emph{{IEEE/ACM} Transactions on Audio, Speech, and
  Language Processing}} \bibinfo{volume}{24}, \bibinfo{number}{5}
  (\bibinfo{year}{2016}), \bibinfo{pages}{927--939}.
\newblock


\bibitem[Simard et~al\mbox{.}(2003)]%
        {SimardSP03_ElasticTransfor_ICDAR}
\bibfield{author}{\bibinfo{person}{Patrice Simard}, \bibinfo{person}{Dave
  Steinkraus}, {and} \bibinfo{person}{John Platt}.}
  \bibinfo{year}{2003}\natexlab{}.
\newblock \showarticletitle{Best Practices for Convolutional Neural Networks}.
\newblock \bibinfo{journal}{\emph{International Conference on Document Analysis
  and Recognition ({ICDAR})}}  \bibinfo{volume}{3} (\bibinfo{year}{2003}),
  \bibinfo{pages}{958--962}.
\newblock


\bibitem[Simonyan and Zisserman(2015)]%
        {SimonyanZ14_VGGStyle_ICLR}
\bibfield{author}{\bibinfo{person}{Karen Simonyan} {and}
  \bibinfo{person}{Andrew Zisserman}.} \bibinfo{year}{2015}\natexlab{}.
\newblock \showarticletitle{Very Deep Convolutional Networks for Large-Scale
  Image Recognition}. In \bibinfo{booktitle}{\emph{International Conference on
  Learning Representations ({ICLR})}}.
\newblock


\bibitem[Sohn(2016)]%
        {Sohn_NPairObjective_NeurIPS}
\bibfield{author}{\bibinfo{person}{Kihyuk Sohn}.}
  \bibinfo{year}{2016}\natexlab{}.
\newblock \showarticletitle{Improved Deep Metric Learning with Multi-class
  N-pair Loss Objective}. In \bibinfo{booktitle}{\emph{Advances in Neural
  Information Processing Systems}}. \bibinfo{pages}{1857--1865}.
\newblock


\bibitem[Takahashi et~al\mbox{.}(2016)]%
        {TakahashiGPG16_CNNDataAugForEventRecog_INTERSPEECH}
\bibfield{author}{\bibinfo{person}{Naoya Takahashi}, \bibinfo{person}{Michael
  Gygli}, \bibinfo{person}{Beat Pfister}, {and} \bibinfo{person}{Luc~Van
  Gool}.} \bibinfo{year}{2016}\natexlab{}.
\newblock \showarticletitle{Deep Convolutional Neural Networks and Data
  Augmentation for Acoustic Event Recognition}. In
  \bibinfo{booktitle}{\emph{Proceedings of the Annual Conference of the
  International Speech Communication Association ({INTERSPEECH})}}.
  \bibinfo{address}{San Francisco, USA}, \bibinfo{pages}{2982--2986}.
\newblock


\bibitem[Tsai(2020)]%
        {Tsai20_LinkingLakhtoIMSLP_Bootleg_ICASSP}
\bibfield{author}{\bibinfo{person}{Timothy~J. Tsai}.}
  \bibinfo{year}{2020}\natexlab{}.
\newblock \showarticletitle{Towards Linking the {L}akh and {IMSLP} Datasets}.
  In \bibinfo{booktitle}{\emph{Proceedings of the {IEEE} International
  Conference on Acoustics, Speech, and Signal Processing ({ICASSP})}}.
  \bibinfo{pages}{546--550}.
\newblock


\bibitem[van~der Wel and Ullrich(2017)]%
        {EelcoU_OMRAugs_ISMIR}
\bibfield{author}{\bibinfo{person}{Eelco van~der Wel} {and}
  \bibinfo{person}{Karen Ullrich}.} \bibinfo{year}{2017}\natexlab{}.
\newblock \showarticletitle{Optical Music Recognition with Convolutional
  Sequence-to-Sequence Models}. In \bibinfo{booktitle}{\emph{Proceedings of the
  International Society for Music Information Retrieval Conference (ISMIR)}}.
  \bibinfo{address}{Suzhou, China}, \bibinfo{pages}{731--737}.
\newblock


\bibitem[Yang et~al\mbox{.}(2019)]%
        {YangTJST19_Sheet2MIDIRetrieval_Bootleg_ISMIR}
\bibfield{author}{\bibinfo{person}{Daniel Yang}, \bibinfo{person}{Thitaree
  Tanprasert}, \bibinfo{person}{Teerapat Jenrungrot}, \bibinfo{person}{Mengyi
  Shan}, {and} \bibinfo{person}{Timothy~J. Tsai}.}
  \bibinfo{year}{2019}\natexlab{}.
\newblock \showarticletitle{{MIDI} Passage Retrieval Using Cell Phone Pictures
  of Sheet Music}. In \bibinfo{booktitle}{\emph{Proceedings of the
  International Society for Music Information Retrieval Conference (ISMIR)}}.
  \bibinfo{pages}{916--923}.
\newblock


\bibitem[Zalkow et~al\mbox{.}(2019)]%
        {ZalkowBM19_SalienceRetrieval_ICASSP}
\bibfield{author}{\bibinfo{person}{Frank Zalkow}, \bibinfo{person}{Stefan
  Balke}, {and} \bibinfo{person}{Meinard M{\"u}ller}.}
  \bibinfo{year}{2019}\natexlab{}.
\newblock \showarticletitle{Evaluating Salience Representations for Cross-Modal
  Retrieval of Western Classical Music Recordings}. In
  \bibinfo{booktitle}{\emph{Proceedings of the {IEEE} International Conference
  on Acoustics, Speech, and Signal Processing ({ICASSP})}}.
  \bibinfo{pages}{331--335}.
\newblock


\end{thebibliography}


\end{document}